\documentclass[12pt,preprint]{aastex}
\usepackage{graphicx}
\def\paren#1{\left( #1 \right)}

\def\be{\begin{equation}}
\def\ee{\end{equation}} 
\def\ltsima{$\; \buildrel < \over \sim \;$}
\def\lsim{\lower.5ex\hbox{\ltsima}}
\def\gtsima{$\; \buildrel > \over \sim \;$}
\def\gsim{\lower.5ex\hbox{\gtsima}}
\def\no{\noindent}

\begin{document}

\title{The Velocity Distribution of Hypervelocity Stars}
\author{Elena M. Rossi$^{1,2}$, Shiho Kobayashi$^{3,2}$ \& Re'em Sari$^{2,4}$} 
\affil{$^1$Leiden Observatory, Leiden University, PO Box 9513 2300 RA Leiden, the Netherlands \\}
\affil{$^2$Racah Institute of Physics, Hebrew University, Jerusalem, Israel, 91904 \\}
\affil{$^3$Astrophysics Research Institute, Liverpool John Moores University, United Kingdom}
\affil{$^4$Theoretical astrophysics 350-17, California Institute of
           Technology, Pasadena, CA, 91125 \\}

\begin{abstract}
We consider the process of stellar binaries tidally disrupted by a
supermassive black hole. For highly eccentric orbits, 
as one star is ejected from the three-body system, the companion remains bound to the black hole.
Hypervelocity stars (HVSs) observed in the Galactic
halo and S-stars observed orbiting the central black hole
may originate from such mechanism.
In this paper, we predict  
the velocity distribution of the ejected stars of a given mass, after they have travelled
out of the Galactic potential. 
We use both analytical methods and Monte Carlo simulations. We find that each part of the velocity distribution encodes
different information. At low velocities $ < 800 $ km s$^{-1}$, the Galactic Potential
shapes universally the observed distribution, which rises towards a peak, related to the Galactic escape velocity.
Beyond the peak, the velocity distribution depends on binary mass and separation distributions.
Finally, the finite star life introduces a break related to their mass.
A qualitative comparison of our models with current observations shows the great potential of HVSs 
to constrain bulge and Galactic properties. 
Standard choices for parameter distributions predict velocities below and above $\sim 800$ km s$^{-1}$ 
with equal probability, while none are observed beyond $\sim 700$ km s$^{-1}$ and the current 
detections 
are more clustered at low velocities $300-400$ km s$^{-1}$. 
These features may indicate that the 
separation 
distribution of binaries that reach the tidal sphere is not flat in logarithmic space, as observed 
in more local massive binaries, 
but has more power towards larger separations, enhancing smaller velocities. In addition, the 
binary formation/evolution 
process or the injection mechanism 
might also induce a cut-off $a_{\rm min} \sim 10 R_\odot$ in the separation distribution.
\end{abstract}


\section{Introduction}

In the last decade, an increasing number of stars were detected,
travelling away from our Galactic center with velocities that exceed
the escape velocity from the Galaxy  \citep{BGK+05,HHO+05,BGK+07,BGK+09,BGK12}.
These stars seem to be consistent with Galactic center origin \citep[see however][but also Brown et al. 2010]{ENH+05}.
These stars, travelling with radial velocities over $\sim 300$ km s$^{-1}$ in the halo,
 are called ``hyper-velocity stars" (thereafter HVSs). 
 Until recently, the discovery technique
automatically selected massive stars, mainly B stars with $M\sim 3-4 M_{\sun}$  \citep[e.g.][]{BGK+07}.
In the last two years, there has been an observational effort to extend the search to an older population (A-stars) of HVS
\citep{BGK+09b,koll+09,koll+10}, but none have been found yet.

HVSs are observed in the Galactic Halo, but they seem to originate from the Galactic center.
Observations of the innermost region ($<0.5$ pc) of our Galactic center
 have revealed young stars in highly eccentric orbits.  
 Between $0.04-0.5$ pc, the stars form one, maybe two, disc-like structures \citep{LB03,lu+09}. 
 At smaller distances, $< 0.01$ pc, instead, they are distributed in a more isotropic fashion.   
These latter are called S-stars and their orbits are the strongest observational 
evidence of the presence of a central black hole (BH) \citep[e.g.][]{genzel+96,ghez+98}
 with mass $M\approx 4 \times 10^{6} M_{\sun}$  \citep[e.g.][]{ghez+08}.
 
 There is here the intriguing possibility that the origin of these stars, 
 and in particular of the S-stars, is related to that of HVSs \citep[e.g][]{PHA07}.
 Theoretically, it has been known for a long time that a tidal field of a central BH
 can break apart the occasional star-binary entering its tidal sphere of
influence, capturing one star and ejecting the other \citep{Hil88}.
The velocity with which a star leaves the black hole potential well -- the {\it ejection velocity
}-- is of the order $ v \approx v_{\rm b}
\left(M/m_{\rm t}\right)^{1/6}$, where $v_{\rm b}$ is the binary orbital velocity, and $m_{\rm t}$ is its total mass \citep{YuT03}.  
Such a velocity, can be
indeed enough ($\ge 1000$ km s$^{-1}$) to climb out of the Galactic potential
and produce a star running through the Galactic Halo
with hundreds  of km s$^{-1}$, as observed. Correspondingly, the semi-major axis of the
captured star $a_{\rm c} \approx G M / (-2 \epsilon) \le 0.02$ pc, where the specific energy is
$\left |\epsilon \right | =v^2/2$.

After the discovery of the first
HVS, there have been a considerable theoretical effort to work out in
more details the consequence of binary tidal disruption
\citep[e.g][]{GPS05,GL06,BKG+06,SHM07,PHA07,KBG+08,
LZY10,SKR10, antonini10, zhang10,zhang13, Kobayashi12}. 
In \cite{SKR10} (hereafter SKR) and \cite{Kobayashi12}, we proposed a new
semi-analytic treatment of the three-body interaction, exploiting the
large disparity in mass $\left(M/m_{\rm t}\right) \gg 1$. This method allows
us to describe a binary-BH encounter independently of most physical
properties of the binary, such as its semi-major axis and the stars'
masses. In particular, we calculated the probability for a binary to
survive disruption and, when disruption occurs, the final energies of
both stars (which is equal but in sign).  Indeed, we showed that the
integration of the binary orbit gives, for the final energy, just a
numerical factor that multiply the analytic expectation (see \S \ref{sec:disruption_model}).
In addition, we found that --- for a parabolic trajectory of the binary center of mass --- 
the ejection probability is independent of the mass of the star, 
yet the heavier component of the binary, if ejected, would escape at a smaller velocity.
Notably, we also showed that the ejection velocity is largely independent of how deep the binary plunges in the 
sphere of influence. In particular, the ejection velocity does {\em not} increase for deeper encounters, as commonly assumed.
 Finally we showed in Kobayashi et al. (2012)
that a parabolic orbit well describe the orbit of 
stellar binaries coming from the edge of the sphere
of influence of SgA*.

 In this paper, we assume a statistical description of the incoming
binaries and calculate the velocity distribution of the ejected stars, using our restricted 3-body method and its results.
This method has never been used before for such a goal, which previously relied on {\em full} 3-body calculations.
The main advantage of our method is that it is computationally inexpensive as well as accurate, 
and the main features of its results can be reproduced analytically. It therefore allows us to investigate different mass and semi-major axis distributions
 for the incoming binaries in a {\em wide range of values}. 
 Moreover, it makes particularly easy to pin down which type of binaries contribute most to a given 
range of ejection 
 velocities.

Exploiting these features of the method, the calculations in this paper are first performed analytically. We assume a
simplified picture of the binary-BH encounter, identify which binaries
(mass ratio and separation) dominate the
production of a HVS with a given velocity and mass. 
These results are then checked against Monte Carlo calculations. The
latter use full numerical integration of binary orbits, and provide a detailed description of the 
velocity distributions.

Qualitatively, tidal disruptions of binaries can yield hypervelocity stars: 
indeed, the maximum possible velocity predicted by SKR ($v_{\rm max} \approx 5000-7000$ km s$^{-1}$) 
substantially {\em exceeds} that seen observationally.
A detailed prediction of the velocity distribution 
and a comparison with current data is therefore required in order to ascertain
 whether the observed distribution is quantitatively consistent with the predictions of a star binary
disruption model. This is the goal of this paper. 
Toward this aim, our findings may be used to better plan future observational 
campaigns. 

The paper is organized as follows. In \S
\ref{sec:disruption_model}, we describe the model we assume for the
binary-BH encounter.  In \S \ref{sec:binary_distributions}, we describe 
our assumptions for the distribution of the binary properties (masses,
semi-major axis, closest approach). A simplified version of our
findings in SKR allows us to derive analytically  the velocity
distributions of HVS (\S \ref{sec:vel_dis_ana}).
Our analytical findings are validated against proper orbit
integrations in \S \ref{sec:montecarlo}, 
where we also perform a comparison with the current sample of unbound HVSs. 
We discuss our results and conclude in \S \ref{sec:conclusion}.

 
\section{Binaries dissolved by black-hole tides}
\label{sec:disruption_model}
We consider a star with mass $m_*$ ejected from a binary system with
semi-major axis $a$ and companion mass $m_{\rm c}$, after a close encounter with a BH of mass
$M \gg m_{\rm t}$, where $m_{\rm t}=m_{\rm c}+m_{*}$.

 We define the penetration factor $D=r_{\rm
  p}/r_{\rm t}$, which measures how deeply into the tidal sphere of
radius \be r_{\rm t} = a \left(M \over m_{\rm t}\right)^{1/3}, \ee a binary
with periapsis $r_{\rm p}$ penetrates. The tidal radius is defined
as the distance from the BH where the mutual gravity of the
binary equals the BH tidal pull. 
When the binary approaches the BH, the orbit of the binary's center of mass has a semi-major axis $r_{\rm a}$ 
comparable to the BH sphere of influence $\sim$ a few pc or $r_{\rm a} > 10^{3} r_{\rm t}$, for $a \lsim 0.1 $ AU. 
The orbital eccentricity is thus $1-e = D (r_{\rm t}/r_{\rm a}) < 10^{-3}$, and the orbit can be 
described with a parabola. In this case, which star is ejected after disruption does not depend on its mass, 
but on its relative position with respect to the BH at the tidal radius
(measured in terms of an angle called ``the binary phase").
Not all binaries, however, are disrupted as a result of the gravitational encounter.
For planar and circular binaries, we showed that
the disruption probability, averaged over the binary phase and sense of rotation, 
is a non-monotonic function of $D$ that does {\em not} saturate to 1 for $D \rightarrow 0$.
Rather,  $\sim 20 \%$ of the binaries survive disruption for $D \ll 1$ (see figure 4 in SKR).
In this paper, we allow for different inclinations and we obtain that 
a smaller percentage, about $\sim 10\%$, survives for $D \rightarrow 0$ (Kobayashi et al. in prep.).

When the binary dissolves, the ejection velocity is
\be
v \cong \sqrt{\frac{2\,G m_{\rm c}}{a}} \left(\frac{M}{m_{\rm t}}\right)^{1/6}.
\label{eq:vedj}
\ee
This is formally the velocity at infinity, in solely presence of the black hole potential. 
Practically,
 the star velocity may be consider constant and equal to eq.\ref{eq:vedj} at a finite distance of $\approx 100 r_{\rm t}$, 
which for binary separation $a \le 60$ AU is smaller than
the sphere of influence of the black hole ($\approx 3$ pc, \cite{Schoedel03}). This range of separations include all the tight binaries that indeed
will produce HVSs. Beyond a few parsecs, deceleration 
from the Galactic potential should be taken into account \citep{KBG+08}.  
In eq.\ref{eq:vedj}, there is a numerical coefficient missing, which depends on the binary phase, its
penetration factor, the binary eccentricity and the inclination of its orbit.
However, the mean velocity over the binary phase is almost
independent of the penetration factor and the coefficient is very close to unity  (see figure 10 in SKR). 

For  our analytical calculations, we thus assume a simple description of the disruption process where
{\em only} and {\em all} binaries approaching the black hole with $D \le
1$ will be tidally disrupted and the final specific energy after breakup is $\pm \epsilon = v^2/2$, where $v$ is given by eq.\ref{eq:vedj}.
 In the Monte Carlo simulations presented later, instead, we compute numerically the final energy for circular binaries, if {\em actually} 
 the binary gets disrupted by the encounter with the BH.
In the Monte Carlo approach, we, thus, drop the two main assumptions of our analytical method: the use of 
eq.\ref{eq:vedj} with a unity proportional factor and of a step function probability distribution, 
where survival occurs only for $D>1$. We will show that our analytical predictions are overall validated, but there are qualitative differences 
in the fastest branches of the velocity distribution, when using a more precise calculation of the disruption energy (Section \ref{sec:montecarlo}).


\section{Binary statistical description}
\label{sec:binary_distributions}
The post-disruption distributions of ejection velocities depend on
the physical properties of the
binaries that reach the BH within its tidal radius ($r_{\rm p}<
r_{\rm t}$).  These are binaries with specific orbital angular momentum $J$
around the BH smaller than $J_{\rm lc}\simeq \sqrt{2 G M r_{\rm
  t}}$. These orbits are called ``loss-cone'' orbits.  The way stars
enter the loss cone depends on the relaxation processes that
redistribute stars in phase space (Lightman \& Shapiro 1977).
In this paper we do not thoroughly investigate the rate of production of HVS. We cite two extreme limits.

If the root mean square
change $\Delta J$ in angular momentum over an orbital period is small, $\Delta J \ll
J_{\rm lc}$, the binary diffuses in phase space until $J\simeq J_{\rm lc}$ and
the binary centre of mass approaches the BH with a flat distribution in $D$. 
In this regime, the rate of disruption of binaries is about one star every orbit at the radius of influence of the BH.
Somewhat counter intuitively, this rate is independent of the binaries tidal disruption radius. We call this regime the 
``empty loss-cone regime". 

In the opposite case of a ``full loss-cone regime" the change in angular momentum over an orbital period is
$\Delta J \gg
J_{\rm lc}$. Then the loss cone is continuously and uniformly
refilled. In this case, the fraction of stars with apocenter $r_{\rm a}$, but with periapsis distance $r_{\rm p} \ll r_{\rm a}$,
is of order $r_{\rm p}/r_{\rm a}$. The rate of production of HVS is therefore proportional to the tidal radius.
Since $r_{\rm t} \propto a$, wider binaries are more frequently disrupted.

Observationally, massive binaries  have
a uniform distribution in the logarithm of the period \citep{DM91,KK12}
which translates into a semi-major axis distribution 
\be
f_{\rm a} \propto \frac{1}{a},
\label{eq:fa}
\ee
where the minimum separation $a_{\rm min}$ should be chosen so to exclude binaries which are 
close enough that they can undergo mass transfer between the binary members \citep{eggleton83}. 
We will discuss the conditions in section 4.1 when the highest velocities are evaluated. 
The upper limit is not very relevant to this investigation as wide binaries produce very low ejection velocities. 
There is observational and theoretical evidence that massive binaries host approximately equal mass stars \citep{KF07, krumholz09,KK12}.
However,  selection effects make it hard to detect binaries with significantly different mass components, and a full observational determination of the distribution of 
binary masses is not available. Here we will consider two extreme cases: i) all binaries are composed of equal mass stars, or ii) the companion mass is drawn
from a power-law mass function,
\be
f_{\rm m} \propto m^{-\alpha}_{\rm c}, 
\label{eq:fm}
\ee
for $m_{\rm min} \le m_{\rm c} \le m_{\rm max}$. 
We define $R_{\rm min}$ and $R_{\rm max}$  the radii of  stars with mass $m_{\rm min}$ and $m_{\rm max}$, 
respectively.
A Salpeter mass function has $\alpha=2.35$. For numerical evaluations, we assume thereafter a 
 low-mass cut-off of $m_{\rm min}=0.5 M_{\sun}$ and a high-mass cut-off of $m_{\rm max} = 100 M_{\sun}$. 
  In our calculations, we will explore the dependence on the index $\alpha$, 
but we will confine ourself to $\alpha >1$, which corresponds to most stars 
being of low mass, as normally observed. 

Finally, we will adopt a linear relation between the radius of the star and
its mass: $R_{\rm c} \propto m_{\rm c}$. This follows from hydrostatic equilibrium and the approximate constancy
of the central temperature for main sequence star. More accurate treatments 
show that $R_{\rm c} \propto m_{\rm c}^{0.8}$ which would not change our conclusions.

\section{Velocity distribution: analytical method}
\label{sec:vel_dis_ana}

If a star with given mass $m_*$ is ejected as a consequence of a binary tidal disruption by a BH of mass $M$, 
what is the probability distribution of its ejection velocity $v$?

For given masses of the HVS and the BH, the ejection velocity is
a function of the binary semi-major axis and the companion mass: $v=v(a,m_{\rm c})$ as given by equation (\ref{eq:vedj}),
\begin{equation}
 v \propto \left\{\begin{array}{lll}
 a^{-1/2}\, m_{\rm c}^{1/2}, & \,\; m_{\rm c} \ll m_*, \\
 a^{-1/2}\, m_{\rm c}^{1/3}, & \,\; m_{\rm c} \gg m_*.
\end{array}\right.
\label{eq:mv_fixa}
\end{equation}
We denote by $\dot{N}_{\rm v}$ the number of stars with velocities of order $v$ ejected from the vicinity of the massive BH per unit time.
In this paper we do not calculate the normalization of $\dot N_{\rm v}$, but
instead focus on its functional shape as a function of $v$.
In principle,  $\dot N_{\rm v}$ should be obtained by integrating over the contribution of all binaries.
However,  the main features of this function can be obtained analytically as follows.
We first calculate how the probability density 
behaves, fixing one parameter at the time: the binary semi-major axis or the companion mass. 
Then, we focus on the binaries that are responsible for the highest ejection velocities and calculate their distribution.
From these results, it is possible to infer which binaries contribute most to a given ejection velocity $v$.
Sketches will visualise the method, complementing the explanation in the text.

The rate, $\dot N_{\rm v}$, at which
stars are ejected from the Galactic black hole with velocity of order $v$ is not, however, the
quantity that we need to compare with observations. In fact, 
we observe HVSs in the Galactic
halo, after they have travelled far from their birth place.
Additional effects, specifically the Galactic potential and the finite age of the stars,
should also be taken into account. We leave that for section \ref{sec:fv_halo_analytics}.

\subsection{The empty loss-cone regime}
\label{sec:empty_dv}

When diffusion is the process that determines how binaries fall into the loss-cone,
the binary approaching rate is almost independent of the size of the loss cone itself \citep{LS77}.
 
If we fix $a$, a mass interval between $m_{\rm c}$ and $m_{\rm c}+dm_{\rm c}$, corresponds to a given 
velocity interval between $v$ and $v+dv$. 
Since the number of ejected stars in these two ranges should be the same, 
the probability density for a {\em fixed semi-major axis $a$}, $\left. f_{\rm v}(v) \right|_{\rm a}$, is such that
$\left. f_{\rm v}(v) \right|_{\rm a} dv  = f_{\rm m}\, dm_{\rm c}$. The relative number 
of ejected stars with velocity $v$ for a fixed $a$ is then $v \times \left. f_{\rm v}(v) \right|_{\rm a}  \propto m_{\rm c}^{-(\alpha-1)}$,

\be
v \times \left. f_{\rm v}(v)\right|_{\rm a}  \propto \left\{\begin{array}{lll}
v^{-2 (\alpha-1)},  & \,\; m_{\rm c} \ll m_*, \\
v^{-3 (\alpha-1)}, &\,\; m_{\rm c} \gg m_*,
\label{eq:fv_ae}
\end{array}\right.
\ee

\no
where we used eq.(\ref{eq:mv_fixa}). In Fig.1, eq.(\ref{eq:fv_ae}) is plotted with thin solid lines, for decreasing separation $a$ towards the right.
The break occurs at the ejection velocity that corresponds to an equal mass binary ($m_{\rm c}=m_*$),
which decreases with $a$ as $a^{-1/2}$ (eq.\ref{eq:mv_fixa}).
The probability density associated with this velocity is simply the probability density for a {\em fixed companion mass},
$\left. f_{\rm v}(v)\right |_{m_{\rm c}},$
evaluated at $m_{\rm c}=m_*$. Since $\left. f_{\rm v}(v)\right |_{m_{\rm c}} dv = f_{\rm a} da$, we get 
\be
v \times \left. f_{\rm v}(v)\right |_{m_{\rm c}} = a \times f_{\rm a} =\, {\rm constant},
\label{eq:vfvm_empty}
\ee

\no
which is constant for our standard choice of  $f_{\rm a}$  (eq.~\ref{eq:fa}).
\begin{figure}
\begin{center}
\epsscale{1.2}
\includegraphics[width=0.85\columnwidth]{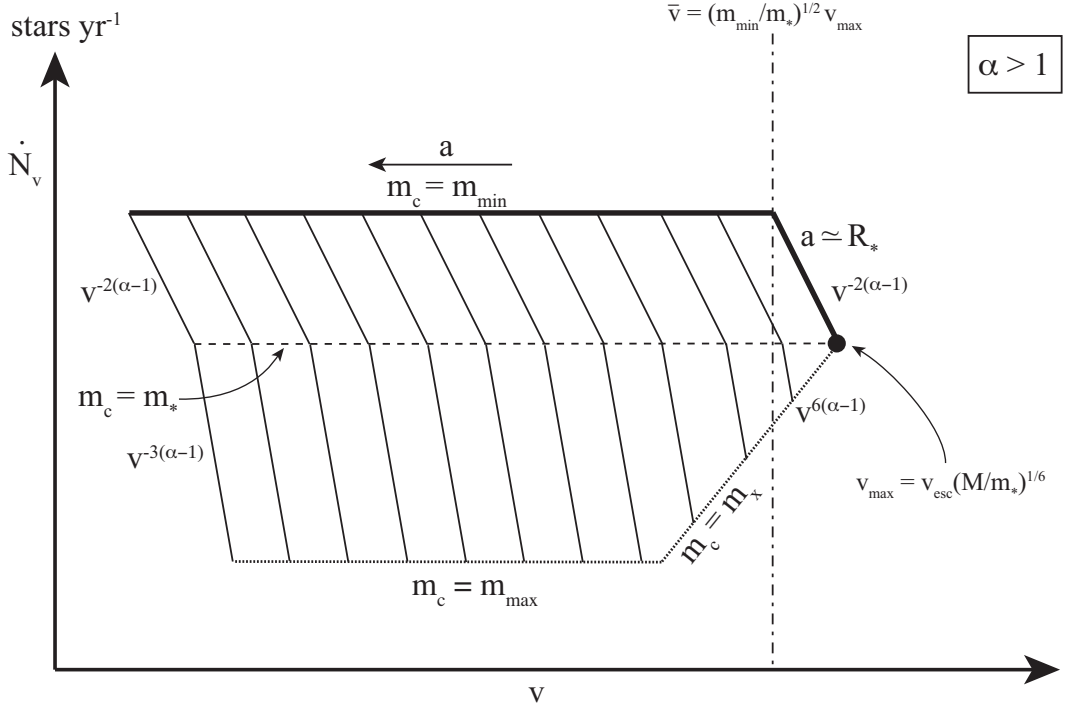}
\caption{Ejection velocity distribution for a given HVS with mass $m_*$. Empty loss-cone regime.
The plot is for $f_{\rm m} \propto m^{-\alpha}$, with $\alpha \ge 1$. 
The solid, thin and broken lines show $\dot{N}_{\rm v}$ for a fixed binary separation $a$ (eq.\ref{eq:fv_ae}).
The rightmost one is for $a=a_{\rm min}$, and the binary separation $a$ increases leftwards.
These lines are equally logarithmically spaced since $ f_a \propto 1/a$ (eq.\ref{eq:fa}).
The dashed line is $\dot{N}_{\rm v}$ for equal mass binaries $m_{\rm c} = m_*$.
The dotted lines mark $\dot{N}_{\rm v}$ for the most massive companion 
allowed for a given binary separation $a$ (eq.~\ref{eq:vxfvx_empty_mx} and eq.~\ref{eq:vxfvx_empty_mmax}).
Finally, the thick black solid line is the total $\dot{N}_{\rm v}$: the low velocity branch is given by binaries with $m_{\rm c} = m_{\rm min}$, while
the high velocity branch corresponds to binaries with $a \approx R_* = a_{\rm min}$  ($m_{\rm c}  \ll m_*$). 
The maximum velocity is obtained for a contact, roughly equal mass binary (the black circle mark). }
\label{fig:empty_sketch_agt1}
\end{center}
\end{figure}

We now specifically consider binaries that produce the fastest ejections. The highest velocities are attained 
for small binary separations and large masses. However, the sum of the stellar radii - which are proportional 
to their masses - can not exceed $a$. In our analytical treatment, if not otherwise stated, 
we will ask a simple relation $a > R_{\rm c}+ R_*$. This implies that for a fixed, small, $a$ the maximum companion mass 
could be smaller than $m_{\rm max}$ and therefore the maximum velocity be less than $v(a, m_{\rm max})$. 

A more realistic model requires a larger minimum separation.
Stars with separations less than a few stellar radii would fill 
their Roche lobes and quickly merge (the exact consequence 
of mass transfer depends on the mass ratio, e.g. Vanbeveren et al. 1998).
Considering that the less massive member has much higher density, 
to avoid such Roche lobe overflow, the separation needs to be larger than
   $\sim R_{\rm c} + R_{\rm c} (m_*/m_{\rm c})^{1/3}$ if $m_{\rm c} > m_*$  
or $\sim R_*       + R_*       (m_{\rm c}/m_*)^{1/3} $ if $m_{\rm c} < m_*$. 
For binaries consisting of very different mass members (e.g. $m_{\rm c} \gg m_*$), the simple constraint on the separation quickly goes to 
$R_{\rm c}$ while in the realistic model it stays more like $2R_{\rm
c}$.  However, these estimates are still approximations since 
the stars change their shape, making it more easy for mass transfer to occur \citep{eggleton83}. 
We will impose $a > 2.5 \max[R_*,R_{\rm c}]$  (e.g. Vanbeveren et al. 1998)
when we numerically discuss velocity distributions in section 6. 
Since $v  \propto a^{-1/2}$, the highest possible velocity is expected
to be reduced by a factor of $\sim 1.6$ or less. As we will see below, the
highest velocity is achieved for $m_*\sim m_{\rm c}$, the correction 
factor for the underestimate of the minimum separation is even 
smaller $\sim \sqrt{2.5/2}\sim 1.1$.

Under the requirement $a > R_{\rm c}+ R_*$, the maximum allowed mass
for the companion, $m_{\rm c,max}$, is the minimum between $m_{\rm max}$ 
and $m_{\rm x} \approx m_* \left(a/R_* \right)$.
The transition between $m_{\rm x}$ and $m_{\rm max}$ happens at  $a \approx R_{\rm max} + R_*$.
For $a < R_{\rm max} + R_*$, binaries with $m_{\rm c}= m_{\rm x}$ have their separation 
nearly equals to the sum of their radii, $a \approx R_*+R_{\rm c}$.
We call them {\em contact binaries}: i.e. binaries which are on the
verge of starting mass transfer. 
They produce the highest ejection velocities for a given separation,

\begin{equation}
 v(m_{\rm x}) \propto m_{\rm x}^{-1/6},   ~{\rm for }\; \, m_{\rm x}  \gg m_*.
 \label{eq:v_x}
\end{equation}

The maximum of all $v(m_{\rm x})$, $v_{\rm max}$, occurs for approximately equal mass binaries: $m_{\rm x} \approx m_*$ and $a \approx R_*$, 
$v_{\rm max} \approx v_{\rm esc} \left(M/m_*\right)^{1/6}$,
where $v_{\rm esc} \approx \sqrt{G m_*/R_*}$ is the escape velocity from the surface of the HVS. 
The result that more massive companions ($m_{\rm c} > m_*$) do not yield higher velocities is a direct consequence of the linear dependence assumed between the radius and the mass of a star. A somewhat shallower relation is obtained combining theory and observations:  $R_* \propto m_*^{0.8}$  (Hansen, Kawaler \&
  Trimble 2004).  Even assuming this latter slope, the conclusion that the maximum velocity is for an equal mass binaries would remain unchanged, since it holds  for any power-law index steeper than $2/3$.

The velocity probability distribution for contact binaries with massive companions, $m_{\rm c} \gg m_*$,
can be derived considering the stars within an area of the parameter space $dm_{\rm c} \times da$, $f_{\rm v}\, dv = f_{\rm a}\, f_{\rm m}\, dm_{\rm c} da$, 
under the condition $a/R_* \simeq m_{\rm c}/m_*$,

\be
v \times \left. f_v \right|_{\rm m_{c}=m_{x}} \propto m_{\rm x}^{1-\alpha} \propto \begin{array}{ll}
             v^{6(\alpha-1)}, & \,\;  m_{\rm x} \gg m_* ,
            \end{array} 
\label{eq:vxfvx_empty_mx}                                                                                
\ee

\no 
where we used eq.(\ref{eq:v_x}),
while for $m_{\rm c} = m_{\rm max} $, 
the relative number of stars is constant (eq. \ref{eq:vfvm_empty})

\be
v \times \left. f_{\rm v}\right |_{\rm m_c = m_{max}} \propto v^{0},
\label{eq:vxfvx_empty_mmax}
\ee
 (see dotted lines in Fig.\ref{fig:empty_sketch_agt1}).

From Figure \ref{fig:empty_sketch_agt1}, it is evident that the total
rate of stars $\dot{N}_{\rm v} \propto v \times f_{\rm v}$ ejected with a given velocity $v$ 
is dominated by binaries with {\em the least possible massive} companion that can give that velocity (thick solid lines),

\be
\dot{N}_{\rm v} \propto \left\{\begin{array}{ll}
            v^{0},   & \,\;  v \ll \tilde{v},\\
            v^{-2 (\alpha-1)}, & \,\; v \gg \tilde{v},
\end{array}\right. 
\label{eq:fv_alphagt2_empty}
\ee
\no 
where $\tilde{v} \sim v_{\rm esc}~(m_{\rm min}/m_*)^{1/2} (M/m_*)^{1/6} = v_{\rm max} (m_{\rm min}/m_*)^{1/2}$.
In (eq.\ref{eq:fv_alphagt2_empty}), the low-velocity tail is
due to ejections from binaries with a low mass companion $m_{\rm c}=m_{\rm min}$ on varying separations, 
while the high velocity branch is given by dissolving contact binaries with $a \simeq R_*$ (i.e. $m_{\rm c} \ll m_*$).
The maximum velocity $v_{\rm max}$ results from dissolving an equal mass binary.
Note that in Figure \ref{fig:empty_sketch_agt1} there is a sharp cut-off at $v_{\rm max}$. 
This is a consequence of our approximate method: a full integration over all binaries that
contribute to a given velocity would result in a more gradual drop to zero of the velocity distribution.


\subsection{The full loss-cone regime}
\label{sec:full_dv}
In the full loss-cone regime the presence of the
BH leaves the star distribution function almost unchanged
and nearly isotropic at all $J$ \citep{LS77}. Contrary to the empty loss-cone regime, this implies a disruption probability as a function of $a$ where
looser binaries are disrupted more easily. More precisely, the cumulative probability to have $D \le 1$ is
\begin{equation}
P_{\rm D \le 1} \propto  r_{\rm t}.
\label{eq:pd1}
\end{equation}

\begin{figure*}
\begin{center}
\epsscale{1.1}
\includegraphics[width=0.85\columnwidth]{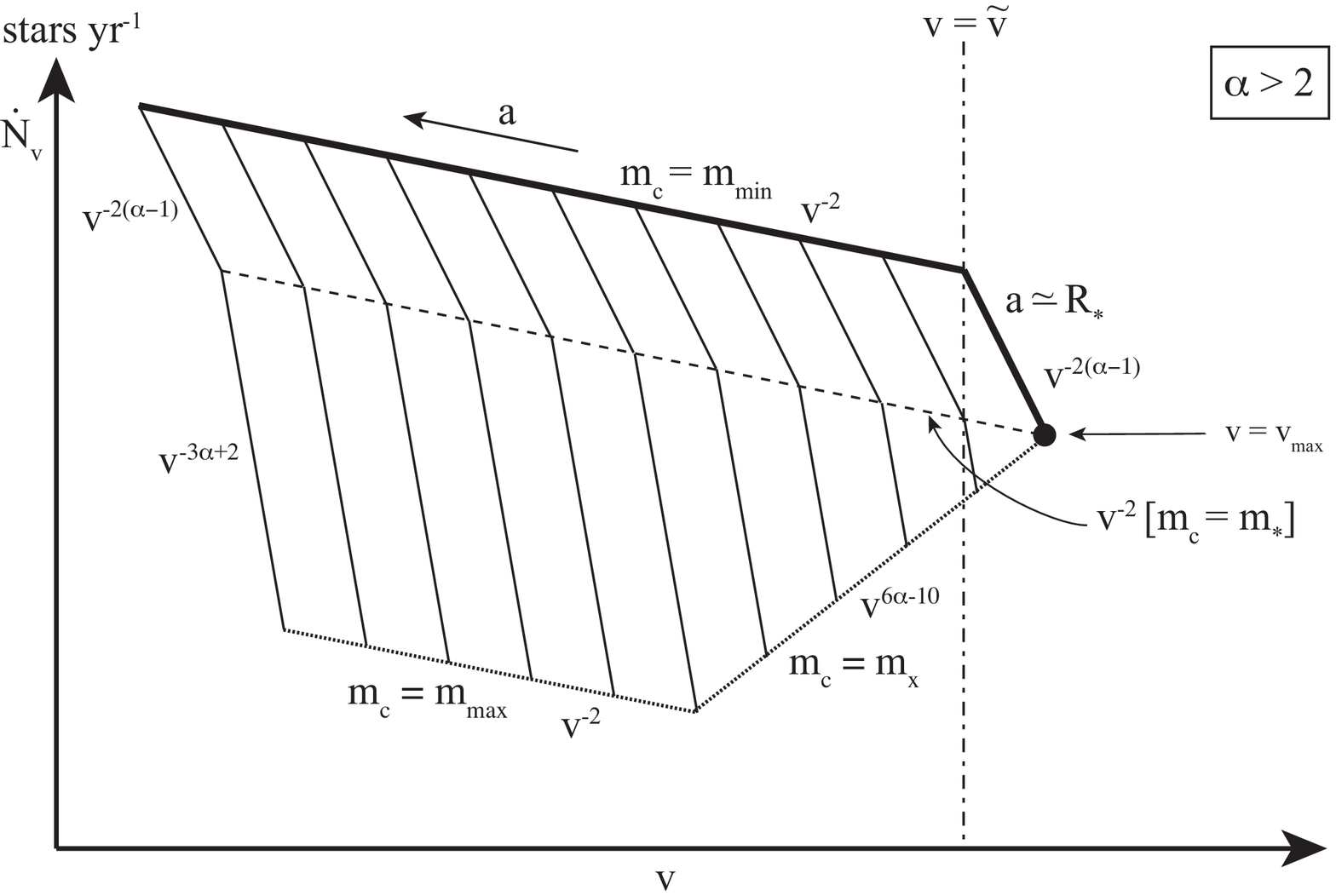}
\includegraphics[width=0.85\columnwidth]{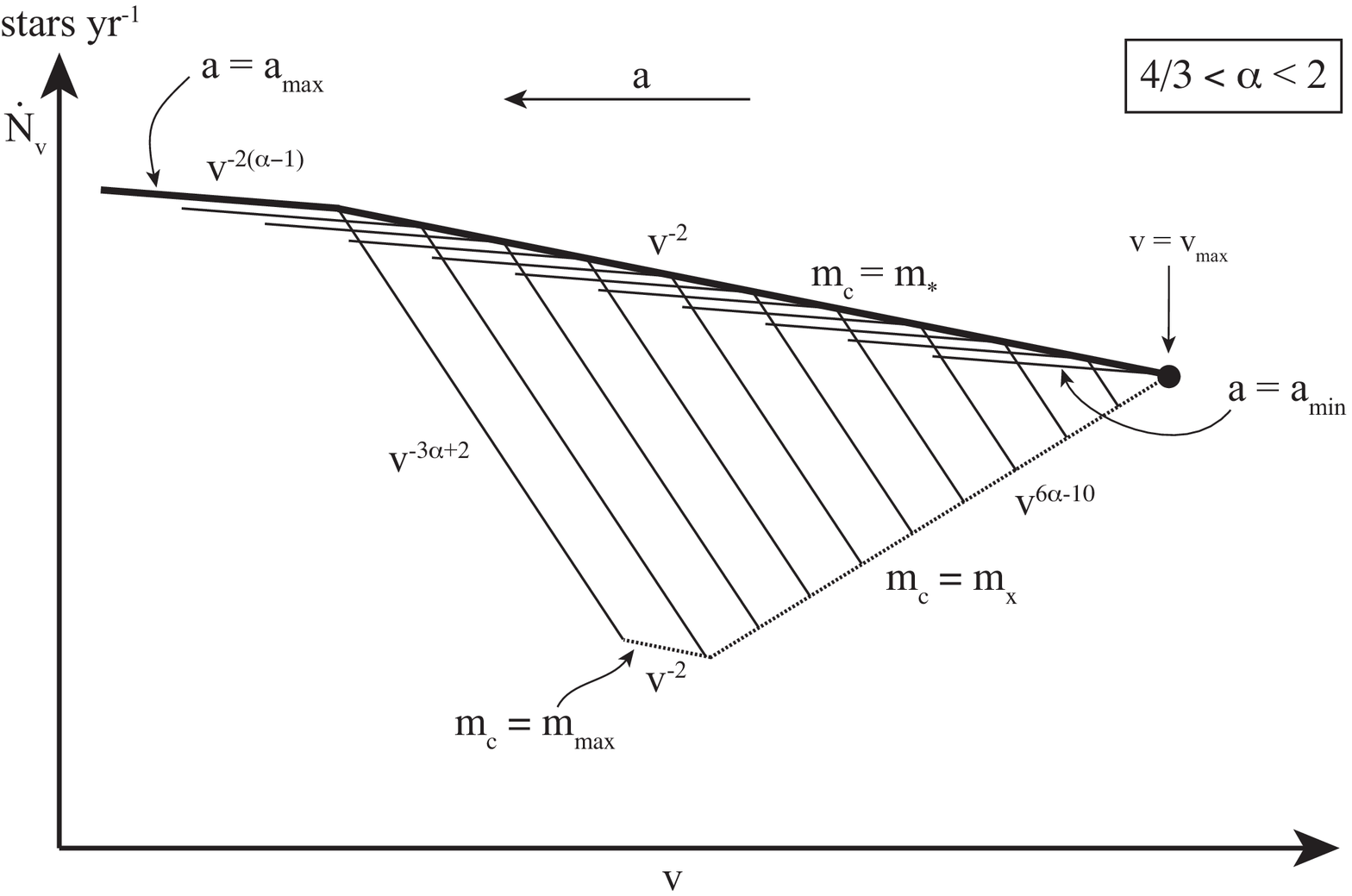}
\caption{The same as Fig.\ref{fig:empty_sketch_agt1}, but for the full loss-cone regime. 
The {\em upper panel} is for $\alpha  \ge 2$. The high-velocity branch is for 
 binaries with  $a \simeq R_*$ and the slope is the same as for the empty loss-cone regime.
The {\em  lower panel} is for $4/3 <\alpha< 2$, and the high velocity branch is populated by 
equal-mass binaries and the slope is shallower than in the upper panel case.}
\label{fig:full_sketch}
\end{center}
\end{figure*}

The velocity probability density for a {\em given} binary semi-major axis is then given by
$\left. f_{\rm v}(v) \right|_a dv = f_{\rm m}(m_{\rm c})\, P_{\rm D \le 1}\, dm_{\rm c}$, where we consider 
that only a fraction $P_{\rm D \le 1}$ of binaries are disrupted.
Therefore, $v \times  \left. f_{\rm v}(v)\right|_a \propto m_{\rm c}^{-\alpha+1} \left(M/m_{\rm t}\right)^{1/3}$, 
\be
v \times \left. f_{\rm v}(v) \right|_a \propto  \left\{\begin{array}{lll}
v^{-2 (\alpha-1)},  &\,\; m_{\rm c} \ll m_*, \\
v^{-3\alpha+2}, &\,\;  m_{\rm c} \gg m_*,
\label{eq:fv_a}
\end{array}\right.
\ee
where we used eq.(\ref{eq:mv_fixa}).
\no
As in the empty loss-cone,  the break occurs at the velocity for an equal mass binary, $v(m_{\rm c}=m_*)$. The distribution is shown in Figure \ref{fig:full_sketch} as thin solid lines that 
corresponds to different separations, which decrease rightwards.
For fixed masses, we have instead,
$\left. f_{\rm v}(v)\right |_{m_{\rm c}} dv = f_{\rm a}\, P_{\rm D \le 1}\, da$, and the relative number of stars 
is given  by
\be
v \times \left. f_{\rm v}(v)\right |_{m_{\rm c}} \propto a \propto v^{-2}.
\label{eq:vfvm}
\ee
An example is shown by the dashed line in Figure \ref{fig:full_sketch} upper panel, for $m_{\rm c} = m_*$.
Finally, we derive the relative number of ejected stars associated with the maximum allowed mass, $m_{\rm c,max}$ for a given separation.
In the case of contact binaries, we follow the reasoning outlined in the previous section, 
taking into account the ejection probability. Since $f_{\rm v}\, dv = f_{\rm a}\,f_{\rm m}\, P_{\rm D \le 1}\, dm_{\rm c} da$, it follows

\be
v \times \left. f_{\rm v}\right |_{\rm m_{\rm c}=m_{x}} \propto m_{\rm x}^{5/3-\alpha} \propto v^{6\alpha-10}, ~~  m_{\rm x} \gg m_*,
\label{eq:vmxfvmx_full_mx}
\ee
where we used eq.(\ref{eq:v_x}).
If the binary is wide enough so that the more massive companion possible is $m_{\rm c}=m_{\rm max}$, then the
velocity distribution is given by (eq.~\ref{eq:vfvm}),
\be
v \times \left. f_{\rm v}\right |_{\rm m_{\rm c}=m_{max}} \propto v^{-2}.
\label{eq:vmxfvmx_full_mmax}
\ee
Eq.(\ref{eq:vmxfvmx_full_mx}) and eq.(\ref{eq:vmxfvmx_full_mmax}) are shown in Fig.\ref{fig:full_sketch} as dotted lines.

Comparing the decay indexes of equations (\ref{eq:fv_a}) and (\ref{eq:vmxfvmx_full_mx}), with $-2$ (eq.\ref{eq:vfvm})
one realizes that there are three different regimes, in which different types of binaries dominate the overall distribution.

For a typical stellar mass distribution with $\alpha > 2$ (see Fig.\ref{fig:full_sketch}, upper panel), 
the main contribution to the total rate of ejected stars with a certain velocity $v$, $\dot{N}_{\rm v}$, comes from binaries
whose companion has the {\em least possible mass} (i.e. the most abundant). 
This is analogous to the case $\alpha >1$, in the empty loss-cone regime, and
\be
\dot{N}_{\rm v} \propto \left\{\begin{array}{ll}
            v^{-2},   &\,\;  v < \tilde{v},\\
            v^{-2 (\alpha-1)}, & \,\; v > \tilde{v},
\end{array}\right. 
\label{eq:fv_full_alphagt2}
\ee
\no
where the low-velocity tail corresponds to binaries with companion mass $m_{\rm c} = m_{\rm min}$
and the high velocity branch is due to contact binaries with $m_{\rm c} \ll m_*$ (thick solid lines in Fig.\ref{fig:full_sketch}, upper panel).

 For a power-law index $4/3<\alpha \le 2$ (see Fig.\ref{fig:full_sketch}, lower  panel)
 the total rate (thick solid lines) is dominated at low velocities by wide binaries (with $a=a_{\rm
  max}$ and $m_{\rm c}\le m_*$), while at high
velocities is dominated by equal-mass binaries ($m_{\rm c}=m_*$),

\be
\dot{N}_{\rm v} \propto \left\{\begin{array}{ll}
            v^{-2(\alpha-1)},    & \,\;  v < v(a_{\rm max},m_*),\\
            v^{-2}, & \,\; v > v(a_{\rm max},m_*).
\end{array}\right. 
\label{eq:fv_full_43alpha2}
\ee

\no
Since $a_{\rm max} \gg a_{\rm min}$, it is in fact likely that $v^{-2}$ is the only
relevant slope for HVSs. 

For a very shallow slope of index $\alpha < 4/3$, most of the power at high velocities 
comes from contact binaries with $m_{\rm c} \gg m_{\rm *}$ and the distribution follows
$\dot{N}_{\rm v} \propto v^{6\alpha-10}$.

The important result here is that the high velocity slope is very steep, with a slope of $-2$
or steeper.  Comparing our results with those of the previous section, we note that in both regimes $ \dot{N}_{\rm v} \propto v^{-2.7}$ for
a Salpeter mass function.


\subsection{Equal mass binaries}

 In case binaries --- especially massive ones --- tend to be of equal mass, ($m_{\rm c}=m_*$), 
the velocity distributions are given by the dashed lines in Figs.\ref {fig:empty_sketch_agt1}
 and \ref{fig:full_sketch}. The relative number of stars in a velocity interval is constant  in the empty loss cone (eq. \ref{eq:vfvm_empty})
and  $\dot{N}_{\rm v} \propto  v^{0}$.
In the full loss cone regime, instead, the relevant distribution is given by eq.(\ref{eq:vfvm}), and the rate of ejected stars goes as 
$\dot{N}_{\rm v} \propto\; v^{-2}$. 
Interestingly, harder binaries have a smaller $J_{\rm lc}$ and could be in the full loss cone regime, while wider binaries may not \citep{gualandris10}.

This would produce a broken power-law distribution from a flat slope to $v^{-2}$, where the break is at the separation $a$ where there is the transition between the two regimes.

\subsection{Summary of our analytical results}

Above we derived the number of ejected stars produced by binaries of all
separations and mass ratios. This allowed us to deduce the total
velocity distribution, the one presented in Figures 1 and 2  by a solid thick line. We here summarize its derivation.

First, we realize that the fastest possible ejection velocity is that obtained from  a nearly 
{\em equal mass, contact}
binary ($m_* \approx m_{\rm c}, a \approx 2.5 R_{\rm *}$),
\begin{equation}
v_{\rm max} \approx 0.56 ~~v_{\rm esc} \left( M \over m_* \right)^{1/6}\approx 3500 \left(m_* \over 3 M_{\sun} \right)^{-1/6} {\rm km \; s^{-1}},
\label{eq:vmax}
\end{equation}
where the numerical estimates are for 
$v_{\rm esc} = (2Gm_*/R_*)^{1/2} \sim 600$ km s$^{-1}$ and $M = 4 \times 10^{6} M_{\sun}$.
If the Roche lobe overflow happens at $a\sim kR_*$, $v_{\rm max}$ would be smaller by 
a factor of $(k/2.5)^{1/2}$.


Lower velocities than $v_{\rm max}$, populating the high velocity branch, can be obtained either by wider equal mass binaries
(e.g. Figure \ref{fig:full_sketch}, lower panel), or 
by binaries with smaller companion masses but 
constant separation $a \approx 2.5 R_*$
(e.g. Figure 2 upper panel), whichever dominates.
In this latter case, the relative frequency of ejected stars is simply given by the higher relative frequency of the lighter companions $\propto m_{\rm c}^{1-\alpha}$. 
Since the ejection velocity depends on lighter companions as $v \propto m_{\rm c}^{1/2}$, the velocity distribution has a slope of $v^{-2(\alpha-1)}$.
On the other hand, the relative frequency of low velocity HVSs that result from wider dissolved binaries depends on the circumstances. 
We have assumed $a \times f_a =$ const., so wider binaries in general are just as common. In the empty loss cone regime, where the rate of dissolving binaries is independent 
of the tidal radius, this leads to a flat distribution $v \times f_v = $ const. However, in the full loss cone regime, where the dissolving rate is proportional
to the tidal radius, this leads to $v \times f_{\rm v} \propto r_{\rm t} \propto a \propto v^{-2}$.

Given a mass function where\footnote{In fact, already for $\alpha>1$ for the empty loss-cone} $\alpha>2$, 
the lighter companion slope for a fixed separation $a$ is steeper than the slope for a fixed mass $m_{\rm c}$ 
(Figure 1 and Figure 2 upper panel). The high velocity distribution is therefore
given by $v \times f_v \propto v^{-2(\alpha-1)}$, down to the velocity
produced by 
a binary with $a \approx 2.5 R_*$
and a minimal mass companion $m_{\rm min}$:
\begin{equation}
\tilde{v} \approx  0.63 ~v_{\rm esc} \left( M \over m_{\rm t} \right)^{1/6} \left( m_{\min} \over m_{\rm *} \right)^{1/2} \approx 1580~ {\rm km \; s^{-1}}.
\label{eq:vtilde}
\end{equation}
The numerical evaluation is for the same values of eq.(\ref{eq:vmax}) and $m_{\min} = 0.5 M_{\sun}$.
Below this velocity, the distribution is either flat, $v \times f_v=$const. (empty loss cone) or 
$v \times f_v\propto v^{-2}$ (full loss cone). 

An implication of our results in both regimes is that, if $a_{\rm min} \approx 2.5 R_*$, 
the high velocity branch in case of $\alpha >2$  depends crucially on the existence 
and physical properties of contact binaries 
(i.e. binaries on the verge of starting mass transfer). 
Observationally, we lack firm constraints. 
It is not clear if contact binaries follow the same mass distribution as wider binaries, 
and if the minimal separation is indeed given by the mass transfer limit or is larger.

\section{Velocity distribution in the Galactic halo}
\label{sec:fv_halo_analytics}

HVSs are observed in the Galactic halo, after they have travelled distances of several tens of kpc.
The velocity distribution at such large distances from the BH can be derived from
$\dot{N}_{\rm v}$, 
once we take into account the effects of the Galactic deceleration and of
the finite lifetime of stars.

The former modifies the low tail velocities, below an effective escape velocity $v_{\rm G}$, where stars consume most of their initial energy to climb out of 
the potential well and reach the halo.
Models for the Galactic potential predict that the escape velocity goes from  $\sim 800$ km s$^{-1}$ at the radius of influence of the BH, to 
$\sim 500-600$ km s$^{-1}$ in the Solar neighborhood\footnote{This is actually observed e.g. Smith et al. (2007).}, 
to $\sim 300-400$ km s$^{-1}$  in the halo at distances smaller than 80 kpc (see model discussion in \cite{KBG+08} and figure 3 in \cite{BGK12}). 
In the following we will assume that
stars are far enough out in the halo, that most of their deceleration has already happened. This is because
most of the deceleration happens just outside the sphere of influence of the BH \citep{KBG+08}, 
while most of the observed HVSs are at distances larger than $\sim 50$ kpc. Since most of the deceleration occurs between a few pc and $\sim 100$ pc, 
an effective Galactic escape velocity could be $v_{\rm G} \approx 800$ km s$^{-1}$.

We can thus relate the velocity at large distances $v_\infty$  to the velocity $v$, which takes only the BH potential into account,
by $v_\infty=\sqrt{v^2-v_G^2}$. 
It then follows that $v_\infty f_{v_\infty}=v f_v  (v_\infty/v)^2$. 
Therefore, for any distribution $f_v$, we obtain a strong low velocity cut-off $ \propto v_\infty^2$ for $v_{\infty} \ll v_{\rm G}$. 
In particular, for power law distributions
$v f_v\propto v^{-\beta}$, we obtain   
\be
v_\infty f_{v_\infty} \propto \left({v^2_\infty}+v_{\rm G}^2 \right)^{-(\beta+2)/2}  v^2_{\infty},
\label{eq:beta}
\ee
which  for ${v_\infty} \ll v_{\rm G}$ recovers the $\propto v^2_{\infty}$ behavior, while for ${v_\infty} \gg v_{\rm G}$, 
has the original power-law slope, which is unaffected by the Galactic deceleration $\propto v^{\beta}_{\infty}$.
We underline that this low velocity break, including the shape around the break, is a robust prediction, since it does {\it not} depend 
on any on the binary distributions nor on the injection mechanism (full versus empty loss cone). In the following,
we will drop the subscript $\infty$ used for convenience to derive eq.\ref{eq:beta}, and $v$ should be understood as the velocity at infinity. 

We are now in the position to compute the most relevant distribution to compare with observations:  the number of stars
with a given velocity $v$ within a distance $r$ from the BH.
At this aim, we need to account for the finite life of stars, $t_* \simeq 10^{10} \left(m_*/M_{\sun}\right)^{-2.5}$ yr \citep{HK94}, and the effect of propagation:
$dN_{\rm v}/dr = \dot{N}_{\rm v} v^{-1}$, which tells us that slower stars are easier to
observe at a given location. From these considerations, the integrated quantity can be calculated as
$N_{\rm v}(<r) = \dot{N}_{\rm v} \times \min[ r/v,t_*]$.
We conclude that the velocity distribution within $r$ has a break determined by both $r$ and the mass of the HVS,
\be
N_{\rm v}(<r) \propto \left\{\begin{array}{ll}       
             \dot{N}_{\rm v}  & \,\;  v <  v_{\rm age}, \\               
             \dot{N}_{\rm v} v^{-1} & \,\; v  > v_{\rm age}.
\end{array}\right. 
\label{eq:fv<r}
\ee 
 HVSs, whose mass is $\sim 3-4 M_{\sun}$, are currently observed out to a distance of  $ \sim 100$ kpc. 
 More massive stars could be in principle observed even at larger radii.
If we are considering distances within $r=100$ kpc and a massive star of $m_* =10 M_{\sun}$ ($m_* = 3 M_{\sun}$), 
the ``age" break in the distribution occurs at $v_{\rm age} = r/t_* \approx 3 \times 10^{3}$ km s$^{-1}$ ($v_{\rm age} = r/t_*  \approx 150 $ km s$^{-1}$). 
Note that we assume here a single age for all stars of a given mass. 
\cite{BGK12b} estimate a 100 Myr time lag between 
formation and ejections of stars in the observed sample. If taken into account, this would just cause a smoothing 
of the break over a very narrow range (a factor $\sim 1.18$) in velocities, and we ignore this correction. 
If the lag time is proportional to the star lifetime, then the correction can be equally ignored for 
any HVS mass.

When $v_{\rm age} \ll v_{\rm G}$ (e.g. for $m_* = 3 M_{\sun}$) the peak of the cumulative 
distribution $N_{\rm v}(<r)$ 
occurs at $v_{\rm peak} = v_{\rm G}/\sqrt{1+\beta}$, while in the opposite case (e.g. for $m_* = 10 M_{\sun}$),
the peak is at $v_{\rm peak} = v_{\rm G}/\sqrt{(\beta/2)}$. For $m_* < 10 M_{\sun}$ and equal mass binaries, 
the value of the index $\beta$ depends on the loss-cone regime and on the binary separation distribution only. With our fiducial choice of $f_{\rm a}$, 
$\beta =0$ in the empty loss-cone, and $\beta = 2$ in the full loss-cone. For more massive stars, $\beta = 2 (\alpha-1)$, 
with a dependence on the mass distribution only.
An accurate description around the peak for $m_*=3 M_{\sun}$ is then given by
\be
N_{\rm v}(<r) \propto \left\{\begin{array}{lll}       
               \frac{v}{v^2+v_{\rm G}^2} & \,\;  v_{\rm peak} = v_{\rm G} & \,\,{\rm for \, empty \, loss \, cone}, \\  
               \frac{v}{(v^2+v_{\rm G}^2)^2} & \,\;  v_{\rm peak} = v_{\rm G}/ \sqrt{3} & \,\,{\rm for \, full \, loss \, cone},            
             \end{array}\right. \nonumber
\label{eq:peak_empty}
\ee 

For $m_* = 10 M_{\sun}$ and a power-law distribution for the companion mass, irrespectively of 
the regime,  
\be
N_{\rm v}(<r) \propto 
			 \frac{v^2}{\left(v^2+v_{\rm G}^2\right)^{\alpha}} \,\; \qquad v_{\rm peak} = v_{\rm G}/\sqrt{\alpha-1} 
\ee 
Finally, for $m_* = 10 M_{\sun}$ and equal mass binaries, we have:
\be
N_{\rm v}(<r) \propto \left\{\begin{array}{lll} 
\frac{v^2}{v^2+v_{\rm G}^2} & \,\;  {\rm no \, clear \, break} & \, \, {\rm for \, empty \, loss \, cone}, \\ 
\frac{v^2}{\left(v^2+v_{\rm G}^2\right)^2} & \, \; v_{\rm peak} = v_{\rm G} 
& \, \, {\rm for \, full \, loss \, cone}.
\end{array}\right. \nonumber
\ee

Beside the peak velocity, equation (\ref{eq:fv<r}) implies other two breaks for unequal mass 
binaries, when $m_* < 10 ~M_{\sun}$: 
at $v_{\rm age}$ and $\tilde{v}$ (ejection velocity imparted when a contact binary contains the lightest 
companion in the distribution, eq.\ref{eq:vtilde}).
For larger HVS mass and for equal mass binaries, instead, there is no break at $\tilde{v}$ and there are only three power-law branches instead of four.
The relative order of these characteristic velocities at a given distance ``r'' depends on the HVS mass. 
For a $m_* = 3 M_{\sun}$, $v_{\rm age} \approx 150 $ km s$^{-1}$ $<v_{\rm peak} <\tilde{v} \approx 1580$ km s$^{-1}$ 
$<v_{\rm max} \approx 3500 ~{\rm km~s^{-1}}$. For $m_* = 10 M_{\sun}$, 
the break ordering is different, $v_{\rm peak} \approx 690 \lsim \tilde{v} \approx 720 ~{\rm km~s^{-1}} < v_{\rm age} \approx 3000 ~{\rm km~s^{-1}}$, 
and a smaller maximum velocity is attained $v_{\rm max} \approx 2900 ~{\rm km~s^{-1}}$.

In the following, we give examples of distributions for $m_* = 3 M_{\sun}$ and $m_* = 10 M_{\sun}$, assuming a Salpeter mass function for
the companion $m_{\rm c}$.
For $m_* = 3 M_{\sun}$, in the {\em empty} loss-cone regime, the distribution is

\be
N_{\rm v}(<r) \propto \left\{\begin{array}{ll}       
                v^{2} & v < v_{\rm age}, \\
                v     & v_{\rm age} < v < v_{\rm peak}, \\
               v^{-1} &  v_{\rm peak} < v < \tilde{v}, \\               
              v^{-3.7} & \tilde{v}<v < v_{\rm max},
\end{array}\right. 
\label{eq:Nr_empty_3m}
\ee 
where $v_{\rm peak}=v_{\rm G} \sim 800 $ km s$^{-1}$.
We note that there is a remarkable steepening after $\tilde{v}$, which makes observations of HVSs with extreme velocities quite unlikely.
For a {\em full} loss-cone, the slope is instead quite steep all the way from the peak, inhibiting 
detections of HVSs with velocity above a few times the peak velocity ($\gsim 10^3$) km s$^{-1}$, 
\be
N_{\rm v}(<r) \propto \left\{\begin{array}{ll} 
                v^{2} & v < v_{\rm age}, \\
                v     & v_{\rm age} < v < v_{\rm peak}, \\
                             v^{-3} & v_{\rm peak} < v < \tilde{v},\\               
                             v^{-3.7} & \tilde{v} <v< v_{\rm max},
\end{array}\right. 
\label{eq:Nr_full_3m}
\ee  
where $v_{\rm peak}=v_{\rm G}/\sqrt{3} \sim 460 $ km s$^{-1}$.
 For $m_* = 10 M_{\sun}$, the velocity distribution is

\be
N_{\rm v}(<r) \propto \left\{\begin{array}{ll}       
                v^{2} & v < v_{\rm peak}, \\
               v^{-2.7} &  v_{\rm peak}<v< v_{\rm age}, \\               
              v^{-3.7} & v_{\rm age} <v < v_{\rm max}.
\end{array}\right. 
\label{eq:Nr_empty_10m}
\ee 
in both loss cone regimes. Here, a sharp steepening occurs right after the peak $v_{\rm peak}=v_{\rm 
G}/\sqrt{1.35} \sim 700 $ km s$^{-1}$.

\section{Monte Carlo calculations}
\label{sec:montecarlo}

We perform Monte Carlo simulations to check whether our analytic 
results can indeed capture the main features of the distributions. 
In particular, we investigate the effect of star-star collisions, and how our assumptions on the energy and
probability distribution affect our results at the high velocity end.  
The large mass ratio $M/m_{\rm t} \gg 1$ allows us to formulate the problem
of a binary- supermassive BH encounter in a restricted three-body approximation, and 
the ejection velocity is given by 
\begin{equation}
v=\sqrt{\frac{2Gm_c}{a}} \paren{\frac{M}{m_t}}^{1/6} \bar{E}^{1/2},
\end{equation}
where the (dimensionless) energy gain $\bar{E}$ at the disruption depends on the geometry of the encounter and
should be computed numerically. In our analytical calculations, we
have assumed a simple model where only and all binaries approaching the
BH with $D\le1$ are disrupted and $\bar{E}=1$. In our Monte Carlo calculations, we instead numerically
integrate the orbit of each realization to determine whether 
the binary survives the encounter and, in case it is disrupted, 
we calculate the actual energy gain $\bar{E}$. Finally --- unlike in the analytic treatment ---
we can take into account the finite size of stars and determine 
whether binary members collide under the tidal force, instead of been torn apart.

\subsection{Procedure} 
\label{sec:procedure}
The initial position and velocity of the binary center of mass are chosen 
so that the binary approaches a massive BH in a parabolic orbit.  
The encounter of a circular binary with the BH is characterized by 
nine parameters: masses of the BH and binary members: $M$, $m_*$ and
$m_{\rm c}$, binary separation $a$, penetration factor $D$, 
binary plane orientation $(\theta,\varphi)$ and binary phase $\phi$ at
the initial distance $r_0$.
As long as a simulation starts at a large enough radius
$r_0\gg r_{\rm t}$, the results are largely independent of it. 
Since we consider HVSs with $m_*=3
M_{\sun}$ or  $10 M_{\sun}$ ejected from a massive BH with 
$M=4\times10^6 M_{\sun}$, we have at most six random variables in 
our Monte Carlo simulations: $m_{\rm c}$, $D$ $\theta$, $\varphi$, $\phi$, and $a$.

For each realization,
we numerically integrate the evolution of the binary 
based on the restricted three-body approximation.
More precisely, we use eqs. 8-10 and 12 in SKR.
The numerical code is provided with a fourth-order Runge-Kutta
integration scheme. We set our initial
conditions at $r_0=10 r_{\rm t}$. There, we assign a penetration factor $D$, an initial binary phase $\phi$  and
the binary plane direction, defined by the two angles $\theta$ and $\varphi$.
There are two special inclinations: prograde binaries, which have their spins aligned with the orbital angular momentum of the center of mass and
retrograde binaries which are instead counter-aligned.

The way we assign $D$ and the various angles is the following. In the empty loss-cone regime, we assume that $D$ is uniformly
distributed between zero and $D_{\rm max}=2.1$ where $D_{max}$ 
is the largest value for which disruption can occur to circular binaries with any binary plane inclination
\footnote{$D_{\rm max}$ occurs for planar
prograde orbits (SKR), which are the easiest configuration to dissolve.}. In fact, for each binary system, there is a different minimum $D$ that a 
 binary can reach without one of the star gets tidally disrupted, $D \sim R/a$, where R is the stellar radius of the more massive star.
 This is generally $\ll 1$, however, it can become of order of $\sim 1$ for contact binaries. 
 Modifications of the high speed branch due to stellar disruptions will be important when/if speeds in excess of 1000 km s$^{-1}$ will be measured.
 Lacking observational motivation, we will omit this effect in this paper.
In the full loss-cone regime, we
randomly choose the pericenter distance $r_{\rm p}$, instead of $D$,
between $0$ and $D_{\rm max} r_{\rm t,max}$ where the maximum tidal
radius is $r_{\rm t,max} = a_{\rm max} \left[M/(m_*+m_{\rm
min})\right]^{1/3}$. This can be translated into the dimensionless distance $D$
once we assign a mass for the companion and a binary separation (see below).
We also randomly choose the binary plane orientation ($\theta$, $\varphi$)
and the binary phase $\phi$
between $0$ and $\pi$. Since a binary starting with a phase difference
$\pi$ has the same post-encounter energy in absolute value but opposite
in sign (SKR), the absolute values of the energies provide the ejection
velocities for the whole range $0<\phi<2\pi$.

If the numerical result indicates that the binary 
survives without disruption or that the minimal star separation
during the evolution is smaller than the sum of the star radii 
(i.e. collision), the realization would be disregarded and another 
one produced, until we have an ejection.
The removal of colliding binaries, prevent us to include spurious very high velocity events 
in the velocity distribution.

The analytical part of the ejection velocity
depends solely on the companion mass and on the binary separation.
We consider two mass distributions for $m_{\rm c}$: the Salpeter mass
function, $f_{\rm m}dm_{\rm c} \propto m_{\rm c}^{-2.35} dm_{\rm c}$ $(0.5\ M_{\sun}
\le m_{\rm c} \le 100 \ M_{\sun})$, and an equal mass distribution: $m_{\rm c}=m_*$. 
The binary separation is assumed to be distributed uniformly in the
logarithmic space: $f_{\rm a} da \propto da/a$ with a maximum separation given by the mean distance of stars in
the sphere of influence of the BH: $a_{\rm max} \sim n_*^{-1/3} \sim
10^6 R_{\sun} \sim 2 \times 10^{-2}$pc  where $n_* \approx 10^5 {\rm
pc}^{-3}$ is the mean stellar number density.  We impose a minimum separation roughly equal to 
the distance at which the Roche overflow will start $a_{\rm min} = 2.5 ~\max[R_*,R_{\rm c}]$.  
However, we do not 
directly use this distribution. To more efficiently evaluate the high velocity tail,
we first adopt a steeper distribution, $f_{\rm a} da \propto a^{-2}da$ $(R_*<a< 10^6 R_{\sun})$.
This function produces high velocity events with higher frequency, since $v\propto a^{-1/2}$.
To reconstruct a velocity distribution that reflects a $f_{\rm a}\propto 1/a$ function,
we then correct our histogram by counting ``a'' times each realisation in a velocity bin, produced by a certain ``a''.
 \begin{figure*}
 \begin{center}
\includegraphics[width=0.85\columnwidth]{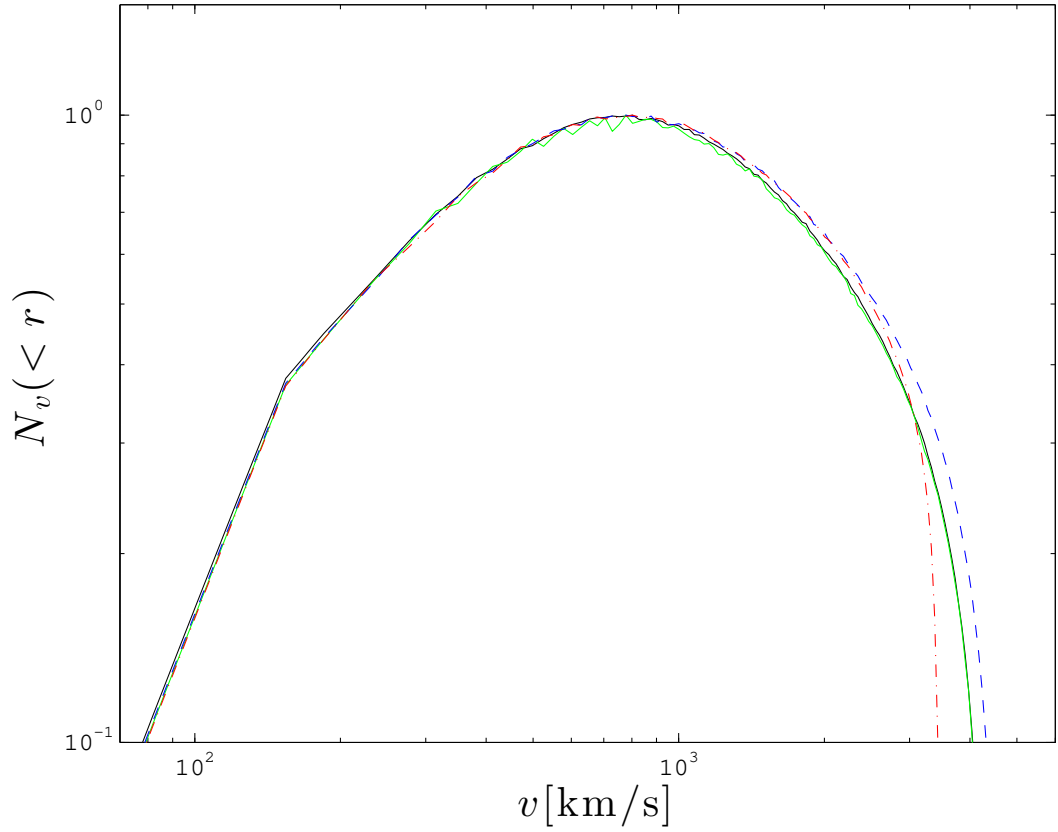}
\caption{A study of our numerical results. We show the case of equal mass binaries with $m_* =3 M_{\sun}$. The solid black and green lines are for 
$10^7$ and $10^{6}$ realisations respectively. The blue dashed line is the
velocity distribution if star-star collisions are not taken into account. Finally, the rad dashed line shows the effect of assuming $\bar{E} =1$.
\label{fig:fig5}}
\end{center}
\end{figure*}

Results in this paper are produced with $10^7$ realisations. A resolution study is shown in
Fig.\ref{fig:fig5}, where the two velocity distributions computed with $10^{7}$ (black solid line) and $3 \times 10^{6}$ (green solid line)
are indistinguishable, especially at high velocities.

\subsection{Results: Velocity distributions}

\begin{figure*}
\begin{center}
\includegraphics[width=0.85\columnwidth]{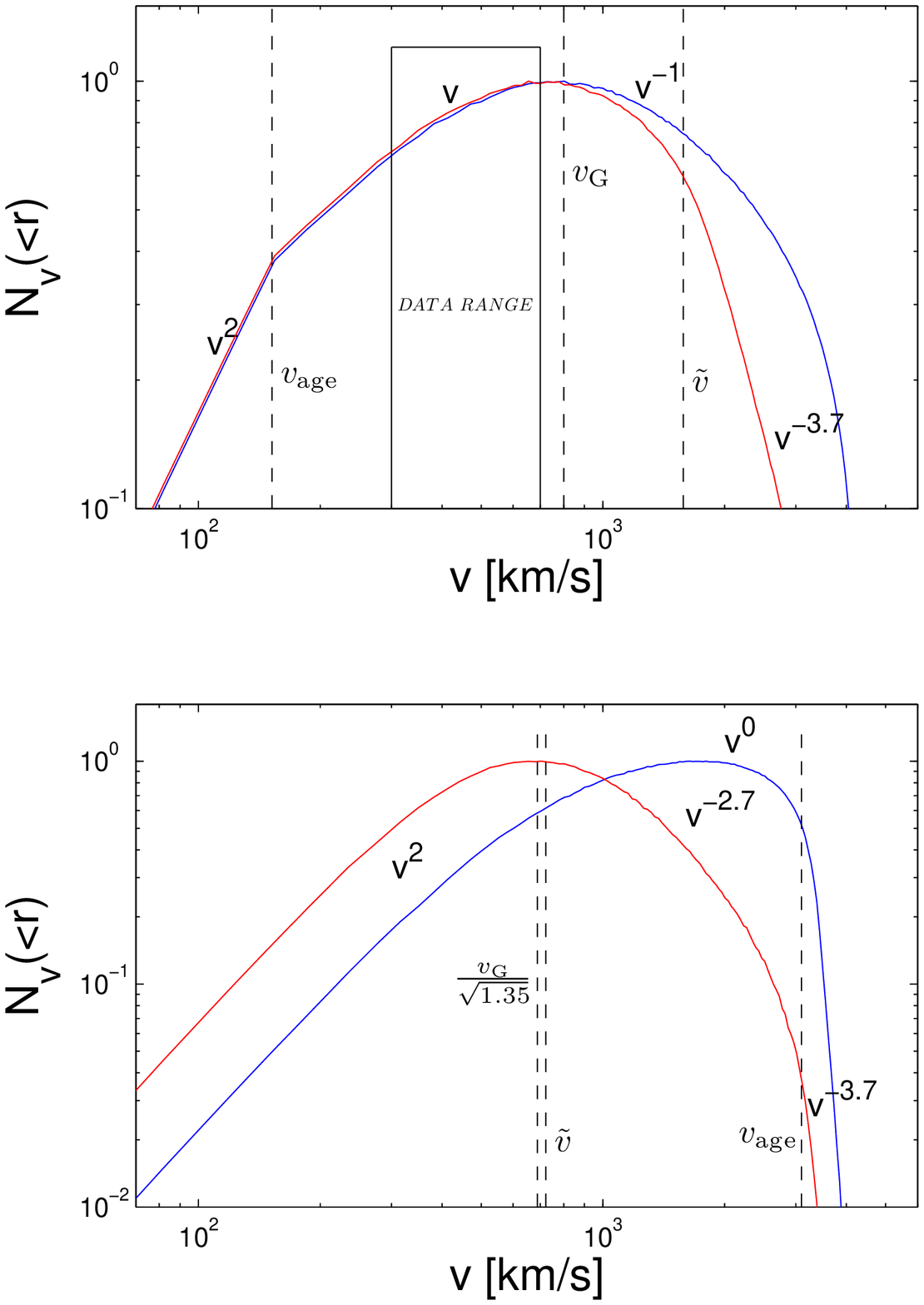}
\caption{Relative number of stars with a given velocity, observed within a radius of 100 kpc: empty loss-cone regime.
In the {\it upper panel} $m_\ast= 3 M_{\sun}$, while in the {\it lower panel}  $m_\ast= 10 M_{\sun}$.
In each panel, the red line is for a Salpeter mass distribution for the companion star and the blue line is for equal mass binaries.
The distributions are all normalized at their peak. Vertical dashed lines mark the characteristic break velocities. 
The marked slopes are the analytical expectations (Sec.\ref{sec:fv_halo_analytics}), 
which well describe our results (see text for discussion). 
Towards $v_{\rm max}$ ($\gsim 3000$ km s$^{-1}$ in the plot),
the slopes gets much steeper than the expected as the distribution will eventually go to zero. 
\label{fig:empty_v}}
\end{center}
\end{figure*}

\begin{figure*}
 \begin{center}
\includegraphics[width=0.85\columnwidth]{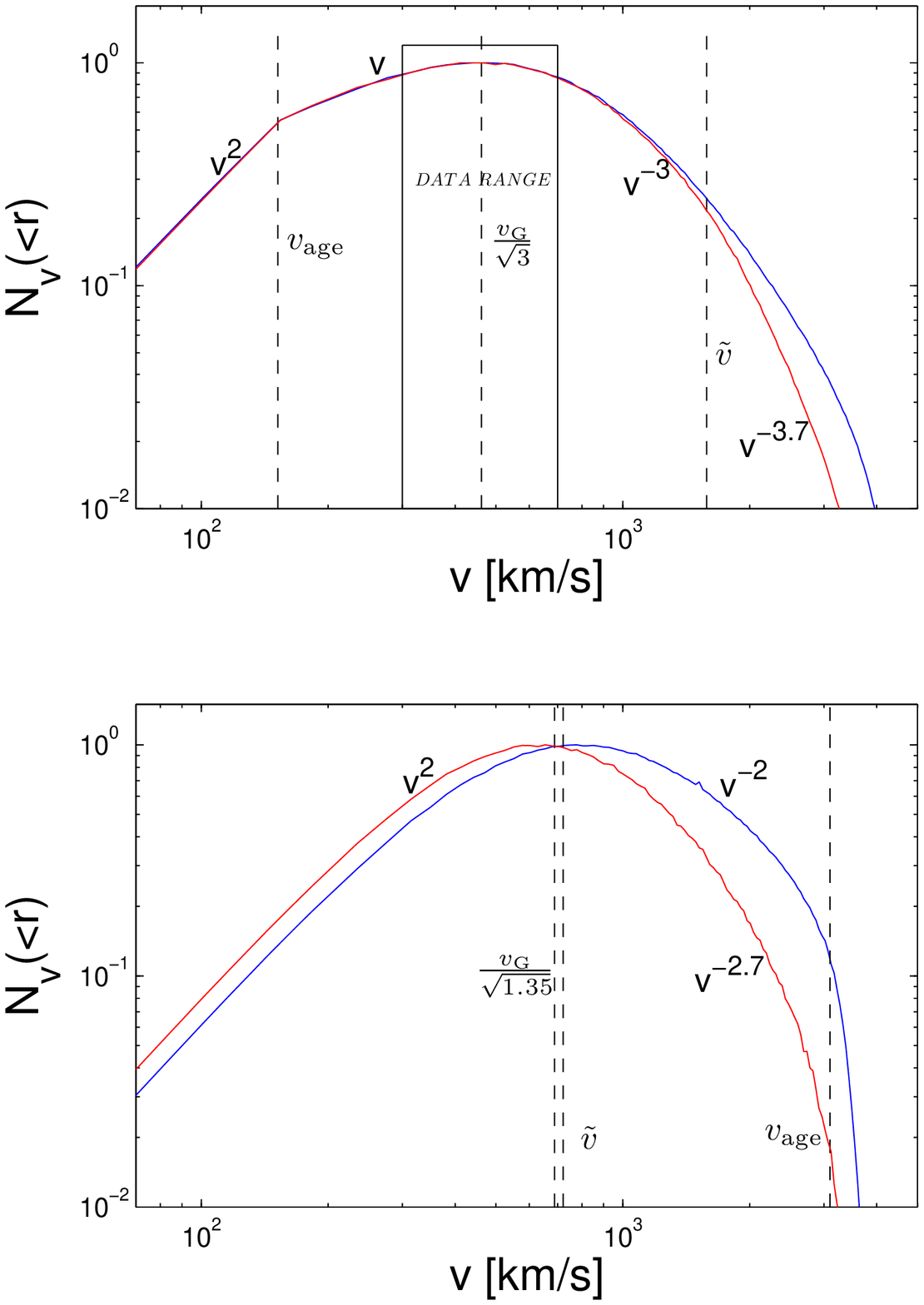}
\caption{The same as Fig.~\ref{fig:empty_v} but for a full loss-cone regime.
\label{fig:full_v}}
\end{center}
\end{figure*}

Our results are shown in Figures \ref{fig:empty_v} and \ref{fig:full_v} for the empty
and full loss-cone regime, respectively. In these Figures, we plot the expected 
velocity distribution within 100 kpc 
\footnote{Distance choices affect only the estimate of the age break $v_{\rm age}$ 
in the figures. For a different distance, the break is shifted
according to $v_{\rm age} \sim 150 ~(r/100 ~{\rm kpc}) (m_*/3M_\odot)^{2.5}$ km s$^{-1}$.}
from the Galactic center for HVSs with $m_\star=3 M_{\sun}$ (upper panels) 
and $m_\star=10 M_{\sun}$ (lower panels).
The distributions with a Salpeter mass function for the companion star are shown with red lines,
while blue lines are for equal mass binaries. 
Our computational method enhances the statistics of the high velocity branches (see Sec.\ref{sec:procedure}) 
and comparatively gives larger statistical errors at low velocities. 
 Furthermore, the Monte Carlo calculations first provide the velocity distribution in terms of the {\em ejection} velocity. When this is converted to the velocity at infinity,  a narrow velocity region just above the escape velocity spans a wide region in the logarithmic space.  To resolve the low velocity tail ($0 < v < v_{\rm G}$), we would need many velocity bins in the numerical calculations and consequently a much larger number of random realizations. However, since  the velocity distribution is expected to be smooth in the narrow velocity region 
 around $v_{\rm G}$, we have used a linear fit to the numerical data to construct the low velocity tail. In contrast, the distribution for velocities  $v> 185$ km s$^{-1}$ comes directly from our numerical results.


Each part of the distribution encodes different information.
The {\em mass distribution} function affects just the highest velocity branch, beyond $\tilde{v}$.
This is because the main contribution to these velocities come 
from binaries with same separation ($a_{\rm min}$) but different companion masses.
In our examples, a Salpeter mass function (red lines) results in a 
significantly steeper slope than for equal mass binaries (blue lines). Shallower power-law slopes than the Salpeter 
one (but greater than 1) would result in intermediate velocity curves, between the blue and the red lines.
The slope of the distribution at the right of the peak (between $\sim v_{\rm G}$ and $\tilde{v}$) is determined by the binary 
{\em separation distribution}, since
the main contribution to this branch comes from binaries with same star masses but different separations. 
As an example, a flatter distribution (e.g. $f_a=$ const.) than a $f_a \propto {\rm 1/a}$
would have more power at larger separation  and therefore it would comparatively enhances lower velocities ($v \propto a^{1/2}$), 
producing a steeper slope in the velocity distribution (we will use this feature when comparing 
our model 
with data in the next section). 
Finally, before the peak, the distribution is determined by the {\em Galactic deceleration}. 

The positions of the breaks bear important information as well. The peak velocity is a measure of the Galactic escape velocity,
the value of $\tilde{v}$ is related to the mass of the lighter possible companion and $v_{\rm age}$ is due to the HVS lifetime. 
However, while the peak velocity may be easy to determine, since we necessarily expect more events there, 
$\tilde{v}$ and $v_{\rm age}$ may be not. 
For example, in our Figures \ref{fig:empty_v} and \ref{fig:full_v}, the break related to $v_{\rm age}$ is below 200 km s$^{-1}$ 
for a $m_\star=3 M_{\sun}$. 
At these velocities, the number HVSs constitute only a tiny fraction of the ``background'' halo stars, whose
Gaussian velocity distribution dominates for $|v|<300$ km s$^{-1}$.
For a $m_\star=10 M_{\sun}$ HVS, the age break occurs 
at very high velocities ($\sim 3000$ km s$^{-1}$), 
where in all cases but one the slopes are very steep, inhibiting detections. 
The break associated with $\tilde{v}$ is again beyond the velocity peak,  
for $m_* \gsim 10 M_{\sun}$ and it is not present at all beyond this mass.

The above features were already understood from our analytical calculations. These latter gives a good qualitative description of the velocity 
distribution (see Section \ref{sec:fv_halo_analytics}) but our numerical results show differences, especially for the high velocity branches beyond the peak.
In our analytical treatment we assumed $\bar{E} =1$ and constant disruption probability for $D<1$.
The mean dimensionless energy distribution shows that mean energies (averaged over all angles) $\langle \bar{E} \rangle $ can instead be larger than unity 
for milder penetrations $0.1 <D<1$. The maximum is $\langle \bar{E} \rangle \approx 1.5$ at $D \approx 0.4$. In this range of $D$, the energy dispersion can also be substantial, 
becoming of order of unity at $D \approx 0.1$. The consequence is that larger ejection velocities can be obtained and the maximum
velocity is larger by a factor of $\sim \sqrt{1.5}$ with respect to our analytical expectations (eq.\ref{eq:vmax}). For instance,
for a $m_*=3$ ($m_*=10$) this implies $v_{\rm max} \approx 4300$ km s$^{-1}$ ($v_{\rm max} 
\approx 3500$ km s$^{-1}$).
On the other, 
the presence of a non negligible dispersion in dimensionless energy causes the slopes to be a bit shallower for the highest velocities, and any break
beyond $\tilde{v}$ to be smoother, than predicted analytically. This is shown, for example, in Fig. \ref{fig:fig5} with the comparison between the solid black (full calculation) and red dashed ($\bar{E}=1$) lines.
Our numerical calculations also show that the average disruption probability (averaged over all inclinations and phases) is quite constant for $D<0.1$, after which it decreases. It is $\approx 0.5$ at $D=1$. This feature makes  ejections with $\bar{E} >1$ less frequent, mitigating the effect described above. In other words, if a constant disruption probability was imposed for all $Ds$, there would be an enhancement of high velocities and the resulting velocity distribution would be, in fact, more different from the analytical expectations at the high velocity end.
Finally, we numerically took into account collisions, when the distance between stars becomes smaller than the sum of the two radii. In Fig.\ref{fig:fig5}, we show that
if not excluded, those realisations yield artificially high ejection velocities (blue dashed line).

\subsection{Comparison with current data}
\label{sec:comparison_data}
\begin{figure*}
\begin{center}
\includegraphics[width=0.85\columnwidth]{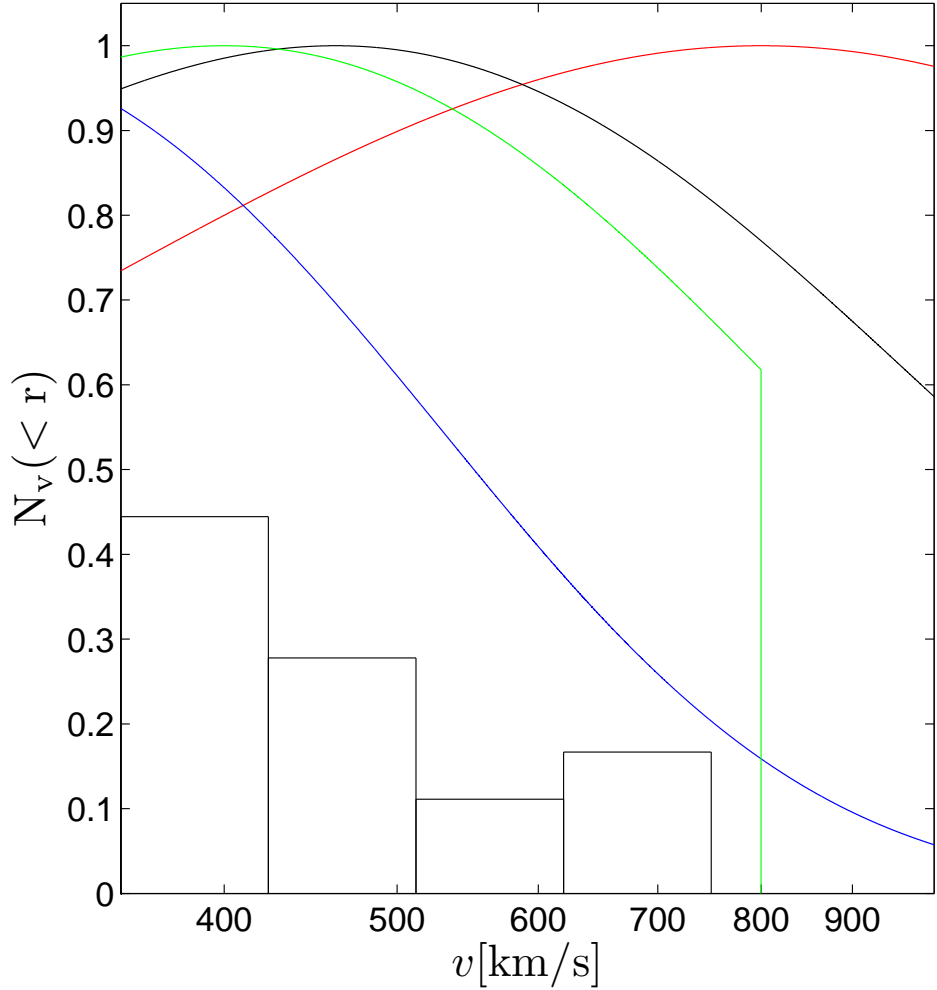}
\caption{Relative number of stars with a given velocity, observed within a radius of 100 kpc for a 3 $M_{\sun}$ star.
Distributions $N_{\rm v} (r<) \propto v f_{\rm v} \propto v/(v^2+v_{\rm G}^2)^{-(\beta+2)/2}$  are normalised at the peak ($= v_{\rm G}/\sqrt{1+\beta}$).
The red and black line are for our fiducial model for an empty ($\beta=0$)
and full ($\beta=2$) loss cone, respectively.  
The green line assume $\beta=3$ and $a_{\rm min} = 10 R_*$ (instead of $a_{\rm min} = R_*$), while the blue line is for $\beta =8$. 
Data in the histogram are taken from Table 1 in Brown et al. (2012). 
The height of each bin indicates the fraction of HVSs in that bin. 
The bins have the same width in the log space.
\label{fig:data_comp}}
\end{center}
\end{figure*}

Current HVS surveys select HVSs with a mass of $3$-$4$ solar masses.
 Up to now, $\sim 18$ HVSs have been observed within $\sim 100$ kpc, 
with radial velocities ranging between $300-700$ km s$^{-1}$ \citep[e.g][]{BGK12}.
Around $300$ km s$^{-1}$, there is an equal number of detected stars which have been classified as {\em bound} HVSs:
they are thought to share the same physical origin as HVSs, but their velocity does not exceed the local Galactic escape speed. 
Below $300$ km s$^{-1}$
the population of stars bound to the halo dominate the velocity distribution and discovery of HVSs using only a velocity component (the radial one)
is inhibited. Beyond ~720 km s$^{-1}$, instead, there are no detections at all, and there is no 
obvious observational bias which explains it.
For our purposes, we will consider only the {\it unbound} sample of $\sim 18$ stars, which has 
a narrow range of observed
velocities ($\sim 380- \sim 720$ km s$^{-1}$) in which the lower end tends to be more populated than the upper end 
(see histogram in Fig.\ref{fig:data_comp}).

We should note that current observations measure only the radial component of the star velocity, 
thus the true three-dimensional velocity can be larger. However, if the observed stars are actually produced in the Galactic Centre,
their trajectories at such large distances --- much larger than the Sun distance from the Galactic Center---
should be almost radial as seen from the Sun. Therefore, we do not expect that a large 
correction should be applied to our predicted velocities, in order to compare our distributions with observations.

\begin{figure*}
\begin{center}
\includegraphics[width=0.85\columnwidth]{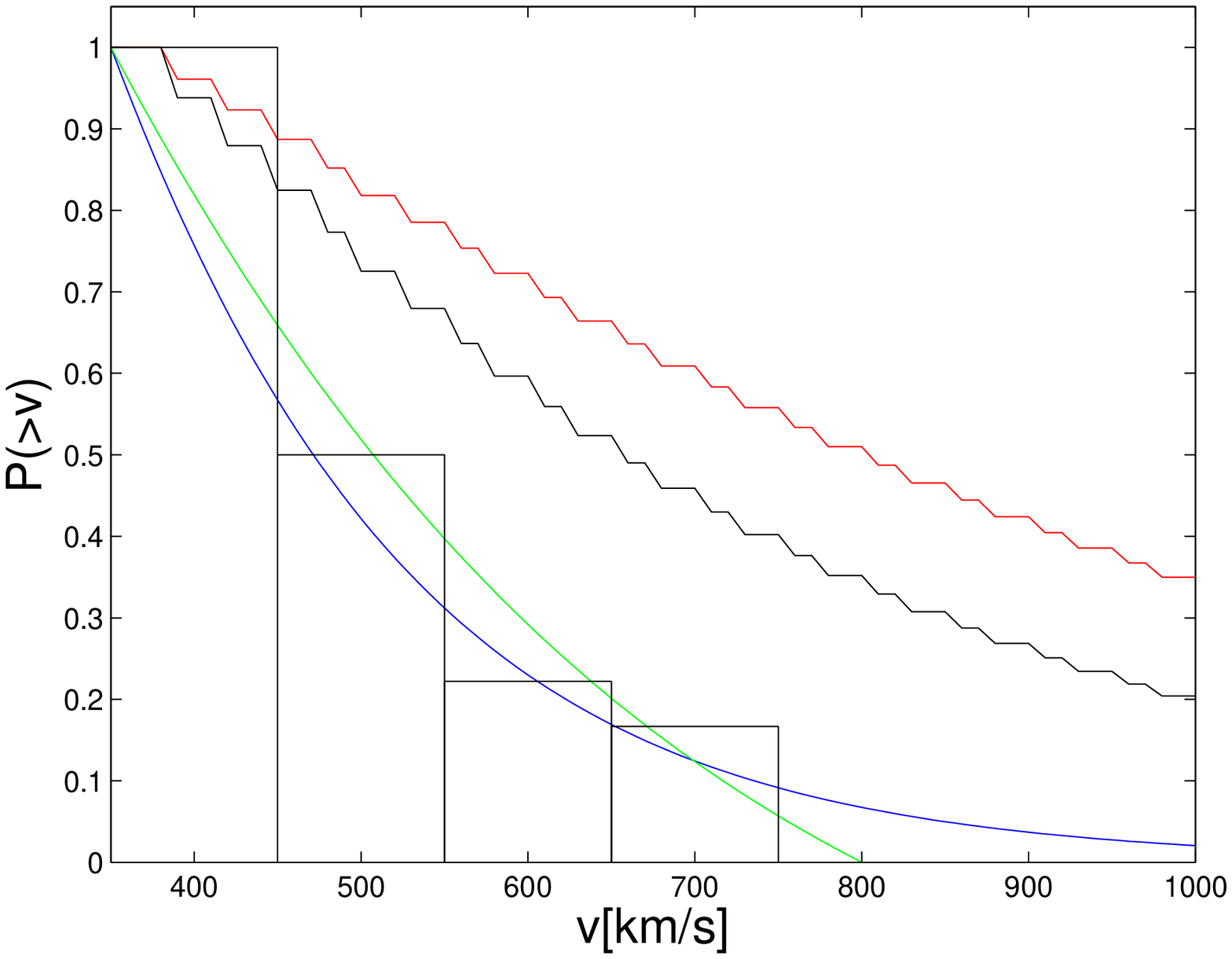}
\caption{The same as Fig.\ref{fig:data_comp}, but for the cumulative distributions. 
This time the fiducial model are plotted using the Monte Carlo data, and not the analytical approximation as in Fig.\ref{fig:data_comp}: they are clearly not consistent with
the cumulative distribution of the current sample. 
The cumulative histogram has the same bin width in the liner space.
\label{fig:data_comp_cum}}
\end{center}
\end{figure*}

Theoretical arguments suggest that our Galactic center favors a situation 
in which the loss cone is empty. {\em Massive} binaries are observed to have periods
 which are flat in logarithmic scale (e.g. Kiminki and Kobulnicky 2012).
 In these conditions, the predicted velocity distribution to compare with data
is shown in  Fig.\ref{fig:data_comp} and Fig.\ref{fig:data_comp_cum} red solid line  (see also Fig.\ref{fig:empty_v}, upper panel). 
Encouragingly, the range of observed velocities lies close to our peak velocity ($v_{\rm G} = 800$ km s$^{-1}$), 
in correspondence with the wide peak of the distribution. However, the data are not {\em including} the peak velocity
(Fig.\ref{fig:data_comp}), while the model would predict that velocities beyond the peak should be as common as those below it. 
In Figures~\ref{fig:data_comp} and \ref{fig:data_comp_cum}, we also plot our fiducial model for a 3 solar mass HVS, 
in the full loss cone regime of star replenishment (black lines, see also Fig.\ref{fig:full_v}, upper panel). 
Although in this case the peak $\approx 460$ km s$^{-1}$ is in the middle of the observed range, the model suffers from the 
same deficiencies, predicting $40\%$ of HVSs with velocity higher than $> 750$  km s$^{-1}$.

We need to add here that observed HVSs may still be decelerating in the Galactic Halo at the observed location, while our model calculate final coasting velocities.
 Using various Galactic potential models from \cite{KBG+08}, we found that a few up to half of the HVSs  can indeed have lower final 
 velocities, but that they remain distributed in the same range $300-700$ km s$^{-1}$. 
 Lower velocities, however, can only strengthen our conclusion, 
that both our {\em fiducial models fail} to account for current observations.

 Analytically, the shape around the peak can be reproduced by the
 function $\propto v/(v^2+v_{\rm G}^2)^{-(\beta+2)/2}$,  which has a
 peak at $v_{\rm G}/(\sqrt{\beta+1})$. In our fiducial models, which
 overpredict high velocities, $\beta=0$ or $\beta=2$. We therefore need
 steeper velocity distribution in the observed range: i.e. larger
 $\beta$ values. We obtain possible $\beta$ values, by comparing our
 model with the observed cumulative distribution
 (Fig.\ref{fig:data_comp_cum}). The blue line is for  $\beta=8$ and the
 green line is for $\beta=3$ plus a sharp cut-off at $v \approx 800$ km
 s$^{-1}$. They both have a probability $\leq 10\%$  for velocity higher
 than $> 750$  km s$^{-1}$. The corresponding differential distributions
 are shown in  Figure~\ref{fig:data_comp}, with the same color
 scheme. The $\beta =8$ curve has a peak around $270$ km s$^{-1}$, fully
 embeded --and therefore hidden -- in the halo stellar distribution. 

What these distribution may mean for our binary/Galaxy physical
parameter? As discussed in Section \ref{sec:fv_halo_analytics}, the
shape around the peak of the distribution for $m_*=3 m_{\sun}$ depends
on three physical ingredients:  
i) the value of the escape velocity, ii) the binary separation
distribution, and iii) the injection mode of binaries into the tidal  
sphere of influence. Since the escape velocity does not influence the
value of $\beta$, we may assume that a steeper slope may be given by a
combination of the last two points. Both an empty and full loss cone
require a binary separation distribution that rises towards larger
separations or at most decreases slowly, to give relative more weight to
lower velocities than to higher ones. In particular, an empty loss cone
requires a binary separation distribution that rises as $f_a \propto
a^{3}$ (for $\beta =8$), while the green line in
Figure~\ref{fig:data_comp} can be reproduced with $f_a \propto a^{1/2}$ 
and a minimum separation for the binaries of $a_{\rm min} \approx 10 R_*
$ (equal to 30 $R_{\sun}$ in this case).  A full loss cone already
favours wide binaries over tight ones, and therefore requires shallower
separation distributions: $f_a \propto a^{2}$ and $f_a \propto
a^{-1/2}$. We may speculate on the origin of these distributions. 
If in the Galactic Central bulge binaries had a separation distribution
that peaks $\gg$ 1 AU (or equivalently period $\gg 30 $ days), then the
shorter period binaries may be indeed described by a rising power-law
distribution in separations. To cite a well known example, binaries in
the solar neighborhood have a log-normal distribution in periods that
peaks at $10^5$ day \cite{DM91}. A possible cut-off at $\sim 10 R_*$ may
instead be related to the binary formation/evolution process or to the
injection mechanism which somehow filters out tight binaries. 

The velocity distributions of HVSs in the Galactic halo have been 
investigated by a few groups through Monte Carlo simulations based on full
three-body calculations (e.g. Bromley et al. 2006, Kenyon et al. 2008,
Zhang et al. 2010, 2013).  Zhang et al. (2010) is especially relevant 
to our study, and they have shown that the velocity distribution 
is sensitive to how binaries move into the BH loss-cone, and 
that if binaries approach the BH in parabolic orbits 
a significant fraction of the resulted HVSs has velocities larger 
than the detected maximum value. They also study the 
interactions between the BH and binary stars bound to in. In this case, 
the binary would experience multiple encounters with the BH. Due to 
the cumulative tidal effect, the probability for the binary to be 
broken up becomes substantial even at a large penetration factor $D$. 
The observed distribution would be reproduced if binary stars 
diffuse onto low angular momentum orbits slowly and most of them 
were broken up at large distance with small ejection velocities. 
This could provide an alternative interesting solution to the problem.


In conclusion, data at this stage seem to suggest the need of more power
to wider binaries, to account for the paucity of high velocities stars. 
The above discussion is of course not conclusive or exhaustive, given
the quality and quantity of the data. Nevertheless, it shows the great
potentiality of HVSs to constrain physical properties of the Galaxy and
the Galactic Bulge, once a larger data set of HVSs will be collected. 

\section{Discussion and Conclusions}
\label{sec:conclusion}
In this paper, we derive the velocity distribution in the Galactic Halo of stars ejected from a stellar binary system, 
following a tidal interaction with the supermassive BH at the center of our Galaxy. 

The magnitude of the ejection velocity, after the star has climbed out 
of the BH potential well, was calculated by SKR and \cite{Kobayashi12}. 
In this paper, we assume a statistical description of the binaries 
injected into the tidal sphere of the black hole (mass and separation distributions)
 and we consider two limiting cases for the probability for binaries to be tidally disrupted:
 one that is linearly proportional to the binary separation (full loss cone regime)
 and one in which all binaries are disrupted with equal probability, (empty loss cone). 
In the former case, looser binaries are disrupted more easily, and since they give rise to lower 
ejection velocities, the velocity distributions in this regime are generally steeper than in the empty loss cone.
A mixed case can be straightforwardly derived from our results, and we use it in our final discussion.
Finally, we account for the deceleration due to the Galactic potential and the finite lifetime of stars,
in order to obtain the star velocity distribution as they travels in the Galactic Halo.  

Some ingredients may be treated less crudely, once the quality of data improve and require more sophisticated models.
An improvement could be to use a detailed model for the Galaxy potential, which includes different Galactic components 
(disc, bulge and halo). It can be a spherically symmetric model \citep{KBG+08} or one that allows for a degree of triaxiality
in the halo \citep{gnedin05}. A larger HVS data sample --- spread over many scales --- may in principle map the shape and depth of the Galactic potential \citep{gnedin05,ym07}.

As discussed in Section \ref{sec:fv_halo_analytics}, one should in principle take a star age distribution, as stars may be injected in the tidal sphere of the black hole at different times in their life. This becomes important when the star travel time to a given Galactocentric distance becomes comparable to the star lifetime, that  occurs for stars more massive than currently observed.
Finally, the highest velocity tail ($\gg $ a few $1000$ km s$^{-1}$), which is given by contact binaries, may be depleted by {\em star} 
tidal disruptions, because the tidal radius for the more massive star of the binary is comparable to the {\em binary} tidal radius. Currently,
these fastest speeds are not observed and therefore we have not included in this paper such modifications. 

Regardless of the above caveats, our results show that, for a given HVS mass, the velocity distributions at large Galactic distances 
have a steep rise up to the peak, which
depends only on the Galactic deceleration model. In our model, where we assume that the stars are far enough 
that all the deceleration has already taken place,
the peak velocity is related to the escape velocity from the Galaxy. 
For velocities larger than the peak, the distribution 
eventually starts decreasing. The slope following the peak depends on the binary 
separation distribution, while the very highest velocity branch bears imprint 
of the binary mass distribution. 

In this paper, we make a first attempt to compare our model with the current data (Section \ref{sec:comparison_data}). 
The observed velocity range is quite narrow (380-720 km s$^{-1}$) and there are no 
detentions of HVSs beyond 720 km s$^{-1}$, though there is no obvious observational bias which explains it. 
By considering  cumulative velocity distributions (fig 7), we find a significant discrepancy between 
the observational data and theoretical distributions based on
conventional assumptions (see also Zhang et al. 2010). 
The detected HVSs are much more concentrated in low velocities around $300-400$ km s$^{-1}$.
To explain the data, we need velocity distributions with index $\beta\sim 8$ much steeper than our 
fiducial values: $\beta=0$ (empty loss-cone) and 2 (full loss-cone). Note that in the empty loss-cone regime 
which theoretical arguments favor, the discrepancy is even larger. If the distribution has a sharp 
cut-off at $v\sim 800$ km s$^{-1}$ (e.g. due to a larger minimum separation 
$a_{\rm min}\sim 10R_\odot$), a shallower distribution ($\beta\sim3$) would be consistent with the data.  
These might indicate that the binary separation distribution in the Galactic 
bulge is not flat in logarithmic space as observed in more local massive
binaries, but has more power towards larger separations, enhancing
smaller velocities (another interesting possibility is the multiple
encounter model studied by Zhang et al. 2010, 2013). 
A possible cut-off at $\sim 10 R_\odot$ might be related to  the binary
formation/evolution process or to the injection mechanism 
which could filter out tight binaries.

At this stage, our model comparison with data is just qualitative, as we need to wait for future missions, 
like the Gaia satellite\footnote{The Gaia mission blasted off in Dec 2013.}, to
detect a few hundreds of HVSs in a wider range of masses. Nevertheless, 
this comparison --- even in its limitation --- shows the great potential of our modeling 
to extract information on the bulge structure and stellar content, and on the Galaxy structure in general.

\end{document}